\documentclass[3p,times,twocolumn]{elsarticle}

\usepackage{graphicx}

\usepackage{amssymb}

\usepackage[figuresright]{rotating}

\newcommand{\jpsi}{J/$\rm\Psi$ }
\newcommand{\psip}{$\rm\Psi$(2S) }

\begin{document}

\begin{frontmatter}

\title{\jpsi and \psip Production
in p-Pb Collisions at 5.02 TeV
with ATLAS}

\author{W. K. Brooks, for the ATLAS Collaboration}

\address{Universidad T\'ecnica Federico Santa Mar\'ia, Avenida Espa\~na 1680 Casilla 110-V, Valpara\'iso, Chile}

\begin{abstract}
The production rates of heavy quarkonia in ion-ion collisions provide sensitive probes in the studies of the hot and dense matter formed in these collisions at high energies. However, a reference for understanding the behavior in the hot medium is necessary; p-A collisions open the possibility to study heavy quarkonia states in a smaller system of much lower average temperature. This is an important step in forming a baseline for understanding A-A collisions, as well as an investigation into the nature of modifications of the parton distributions in the nucleus. Using data collected at the LHC in 2013, we show results on the prompt \jpsi and \psip nuclear modification factors and the double ratio, \psip divided by \jpsi in p-Pb divided by the same in p-p, in p-Pb collisions at 5.02 TeV. The charmonia states were reconstructed via the dimuon decay channel and the yield is analyzed differentially in bins of transverse momentum, rapidity, and event activity.
\end{abstract}


\end{frontmatter}

\section{Introduction}
\label{intro}
\vspace{-0.1cm}

The suppression of quarkonia in heavy nuclear systems has been a topic of study for four decades.
The production rates of \jpsi and \psip mesons from heavy nuclei have been measured and compared in the highest energy fixed-target and collider experiments since 1990~\cite{Hu2015}.  
\begin{figure}[b!]
  \includegraphics[width=8cm,clip=true,trim=1.5cm 0.3cm 0 0]{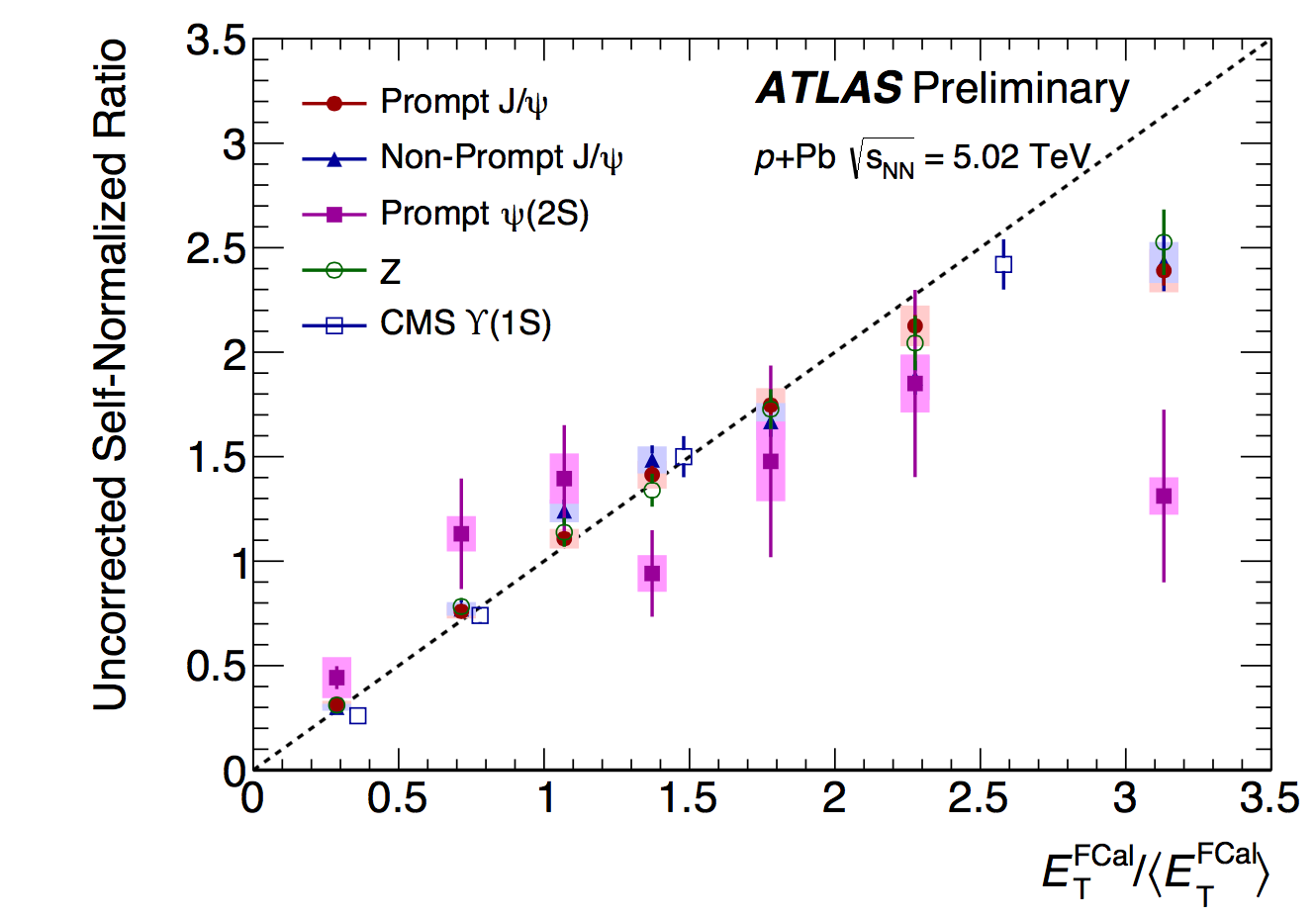}
 \vspace{-0.5cm}
  \caption{The self-normalized ratios for the five measurements from ATLAS and CMS listed in the figure key vs. the self-normalized ratios of forward calorimeters in each respective experiment. The data have not been corrected for any centrality bias, see text.}
  \label{fig:snratios}
\end{figure}

\begin{figure*}[t!]
  \includegraphics[width=\linewidth]{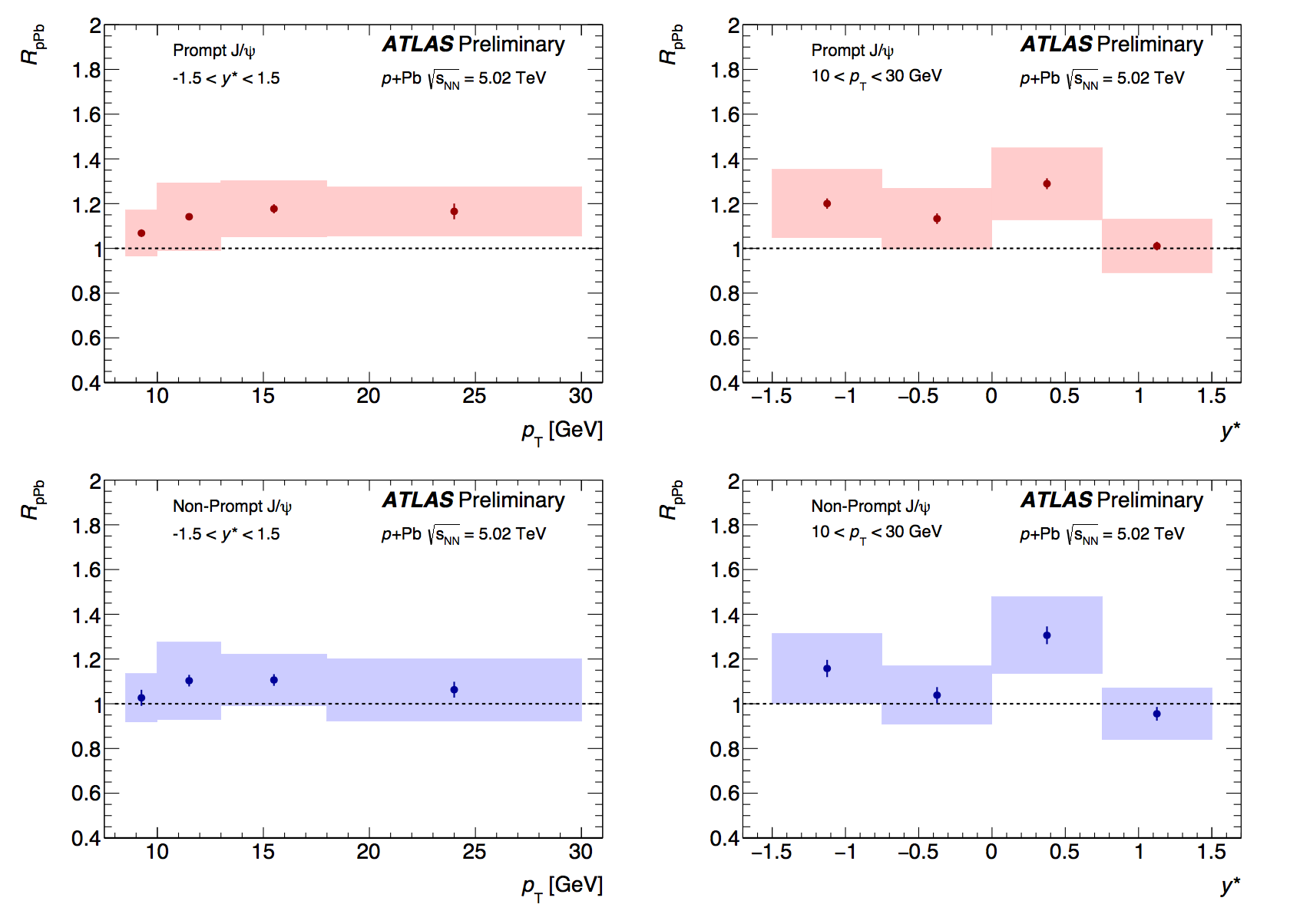}
  \caption{Plots showing the nuclear modification factor $\rm R_{pPb}$ of \jpsi meson production in p-Pb collisions compared to those in p-p collisions. The upper two plots are for prompt \jpsi mesons while the lower two plots are for non-prompt \jpsi mesons. The left column shows the $\rm p_T$ dependence while the right column shows the $\rm y^*$ dependence.}
  \label{fig:rppb}
\end{figure*}
There are several motivations for study of the production of quarkonia in nuclei. While measurements in proton-proton collisions have already provided significant constraints on the production mechanism for these systems, additional information may emerge from their interactions in the extended medium available in proton-ion collisions due to the much higher effective gluon density made available in such systems. Additional information on the formation time and dipole-nucleon or hadron-nucleon interaction cross sections for these mesons can be obtained from understanding the energy dependence of the observed absorption. Further, while the production rates of heavy quarkonia states in ion-ion collisions may provide a sensitive probe of the hot and dense system formed in such collisions, a reference measurement is needed in the cold medium in order to understand the behavior in the hot medium; p-A collisions open the possibility of studying heavy quarkonia states in a smaller system of much lower average temperature. This is an important step in forming a baseline for understanding ion-ion collisions, as well as an investigation into the nature of modifications of the parton distributions in the nucleus. 

ATLAS~\cite{ATLAS_paper} has recently measured the production of the forward-backward ratio of \jpsi production in proton-lead collisions~\cite{Arratia2015} and compared \jpsi and \psip production in these collisions to the rates in p-p collisions~\cite{Hu2015}. The latter comparisons are the focus of the present discussion. 

\section{Experimental Methods}
\label{meth}
\vspace{-0.1cm}

The measurement of the cross sections for \jpsi and \psip in ATLAS was carried out using the dimuon decay channel. One or more triggered muons in the hardware-based first level of the trigger (L1), and two muons found in a full scan of the software-based Event Filter (EF), were required to trigger the event. In offline reconstruction, unlike-sign dimuon pairs were identified as quarkonium candidates. A calculation of the invariant mass and the pseudo-proper lifetime was performed for each candidate in the pair range 8.5~$\rm<p_T<$30 GeV and center of mass rapidity $\rm| y^* |<1.94$. A simultaneous weighted fit to the invariant mass and the pseudo-proper lifetime was performed using a fit model for these lineshapes. The weight of each event consisted of the inverse of the product of the efficiency of the L1 trigger for each muon in the pair, the efficiency of the dimuon trigger at the EF, the reconstruction efficiency for the pair, and the experimental acceptance in the kinematics of each pair.

The collision vertex is determined using the non-muonic charged particle tracks in each event. \jpsi and \psip mesons originating from the decay of b quarks can have a detached vertex reflected in a larger value of  the pseudo-proper lifetime, while directly produced mesons do not have a detached vertex. This is taken into account in the fit using separate "non-prompt decay" and "prompt decay" fitting terms. The fit model consisted of four signal terms and three background terms. Each signal term consisted of the weighted sum of a crystal ball function and a Gaussian function in mass multiplied by a term depending on pseudo-proper time. For the latter term, the prompt quarkonium production was modeled by a delta function while the non-prompt quarkonium production was modeled by an exponential decay. This model was convoluted by a double Gaussian function in pseudo-proper time to take into account the experimental detached vertex resolution. The prompt background in invariant mass was adequately represented as a constant and as a delta function in pseudo-proper time, while the non-prompt backgrounds were modeled by two exponential functions in both mass and pseudo-proper time. Variations on this model were used to estimate the systematic uncertainties, which constituted the largest systematic uncertainties of the measurement, ranging from 2\% to 10\% for \jpsi and from 5\% to 80\% for \psip, depending on the kinematic bin.

Two different datasets were analysed. First, p-p collisions at $\rm \sqrt{s}$ = 2.76 TeV were analyzed. Second, p-Pb collisions at $\rm \sqrt{s_{NN}}$ = 5.02 TeV were analyzed. In order to compare the yields in the p-Pb collisions to p-p collisions, it was necessary to perform an interpolation of the cross sections between $\rm \sqrt{s}$ = 2.76 TeV and $\rm \sqrt{s}$ = 7 TeV, also including data at $\rm \sqrt{s}$ = 8 TeV when it was available, to have the equivalent cross section for p-p at $\rm \sqrt{s_{NN}}$ = 5.02 TeV. The systematic uncertainties of this interpolation were carefully studied and are included in the overall uncertainties as appropriate. 

An issue of high interest is whether there is any measurable centrality dependence of any observable. For central collisions the path length through the medium is longest, potentially accentuating any medium-related modifications. The experimental determination of the centrality in these collisions is more ambiguous than in ion-ion collisions and therefore a number of centrality measures have been explored. An assumption underlying the centrality measures discussed here is that the centrality is monotonically related to the energy deposited in the forward calorimeter (FCal). For more details, see Ref ~\cite{Hu2015}.
\begin{figure}[t]
  \includegraphics[width=8cm,clip=true,trim=0.4cm 0.3cm 0 0]{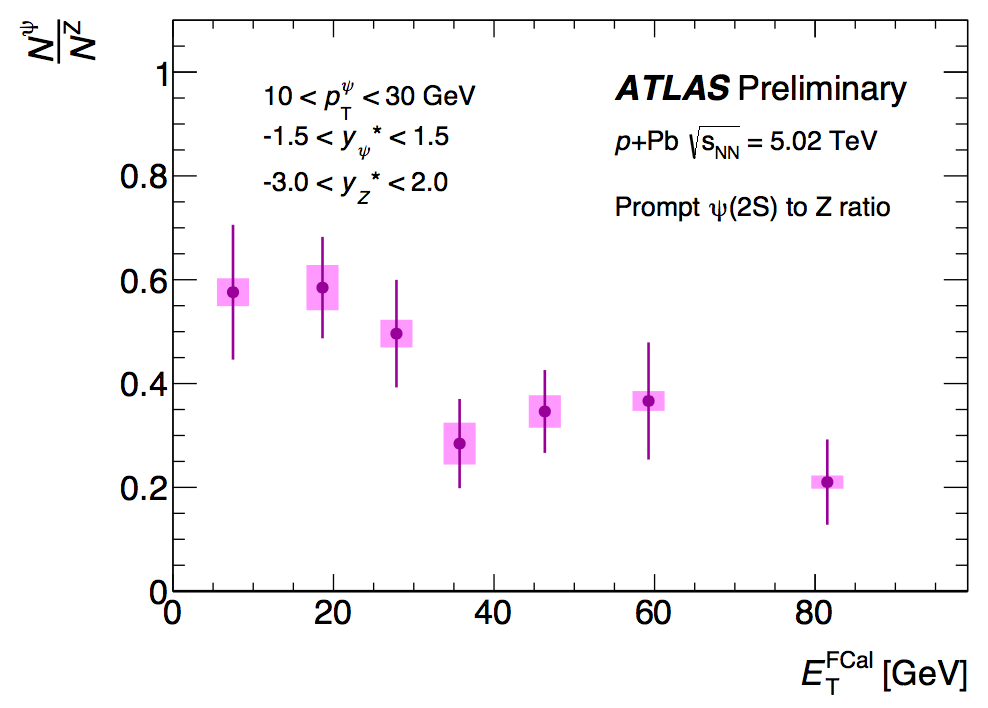}
 \vspace{-0.7cm}
  \caption{Ratio of the yield of \psip to the yield of Z bosons as a function of transverse energy in the forward calorimeter FCal. This method of normalizing the yield is expected to strongly reduce model assumptions about the event centrality measure. The decreasing trend seen suggests a centrality dependence of the \psip production, indicating a dependence on the nuclear medium thickness.}
  \label{fig:psipZ}
\end{figure}

\begin{figure*}[]
  \includegraphics[width=15.0cm,clip=true,trim=0.4cm 0.3cm 0 0]{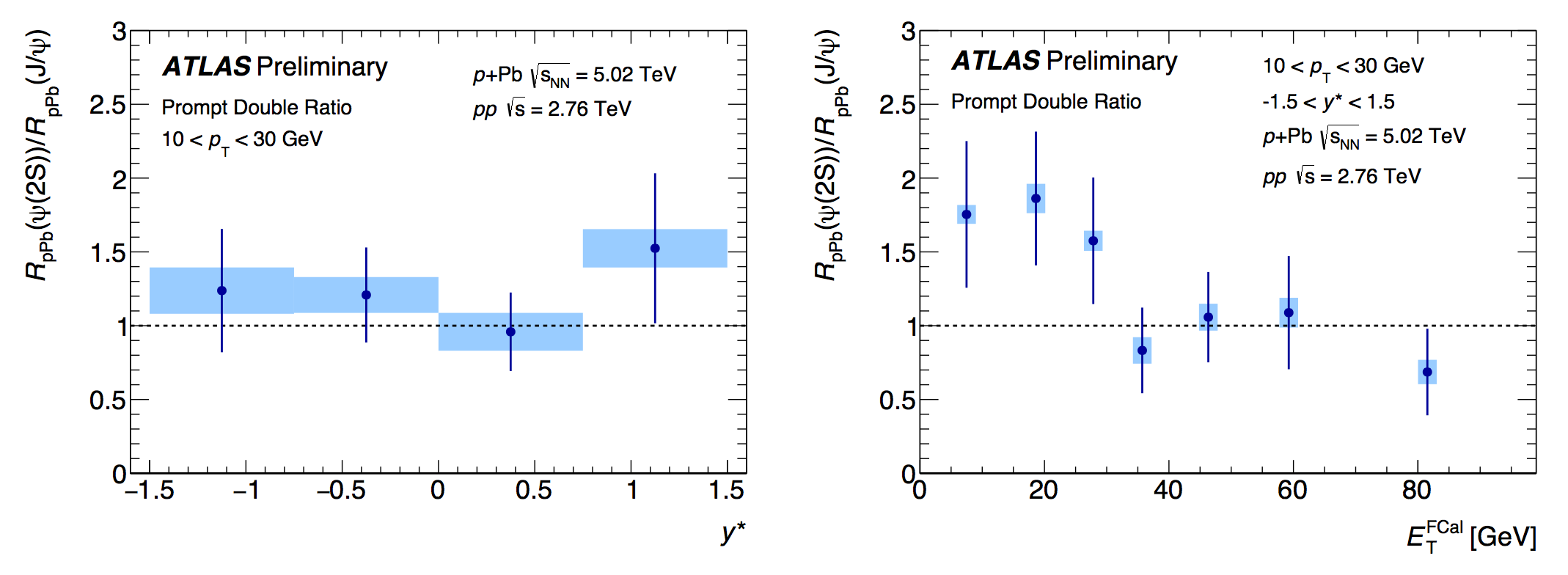}
  \caption{The double ratio $\rm R_{pPb}( \Psi ( 2S ) ) /R_{pPb}( J/\Psi)$ as a function of FCal $\rm E_T$. In the absence of medium effects, this ratio should be equal to unity. A downward trend with increasing FCal $\rm E_T$ is seen, beginning with a value greater than unity.}
  \label{fig:doubleratio}
\end{figure*}

\section{Results}
\label{results}
\vspace{-0.1cm}

An approach to reducing the model dependence of centrality studies is to use the 'self-normalized yields' as recently employed by CMS in studies of the $\rm \Upsilon$(1S) state~\cite{CMS_Upsilon}. Fig.~\ref{fig:snratios} shows self-normalized ratios for prompt \jpsi and \psip, for non-prompt \jpsi, for Z bosons, and the CMS points for $\rm\Upsilon$(1S). As can be seen, for the highest $\rm E_T$ values presumably corresponding to most-central events, all ATLAS measurements fall below the diagonal line, and the \psip is far below the other points. Plots of this type, corrected for centrality bias in two different  scenarios, are given in ~\cite{Hu2015}.

The nuclear modification factor $\rm R_{pPb}$ can be defined as the ratio of the yield in p-Pb to the yield in p-p multiplied by 208. This quantity is shown in Fig.~\ref{fig:rppb} for prompt and non-prompt \jpsi mesons. Generally the trend is for this ratio to be somewhat greater than unity, i.e., the production is enhanced, and there is no strong dependence on $\rm p_T$ or $\rm y^*$. Since for most points the deviation from unity is similar in magnitude to the systematic uncertainty for each individual point, and since the uncertainties are substantially correlated, it is difficult to make strong statements about this enhancement, however, it is an intriguing trend that should be validated by other experiments and future studies. 

Another approach to reducing model uncertainties in centrality assignments is to normalize the quarkonia yield to the yield of particles that do not strongly interact with the medium, such as the vector bosons. In Fig.~\ref{fig:psipZ} is seen the result of this approach when employing Z bosons as a normalization for \psip mesons. The overall trend of this ratio is to decrease with increasing FCal $\rm E_T$. Plots of this kind for prompt and non-prompt \jpsi mesons can be found in Ref.~\cite{Hu2015}, and for those mesons the trend in the ratio is consistent with being independent of FCal $\rm E_T$.

Finally, the double ratio $\rm R_{pPb}( \Psi ( 2S ) ) /R_{pPb}( J/\Psi)$ takes the strongest advantage of cancellation of systematic uncertainties, while the normalization should produce unity in the limit of no medium effects. As can be seen in Fig.~\ref{fig:doubleratio} in the left panel, the mean value of this ratio as a function of $y^*$ is greater than unity, and in the right panel it can be seen that with increasing FCal $\rm E_T$ the trend is a decreasing one, beginning at a value above unity and ending at a value equal to or less than unity. Thus, these data appear to be consistent with an enhancement for peripheral events in \psip production relative to \jpsi production, with a transition toward a possible suppression for more-central events.

\section{Conclusions}
\label{conclusions}
\vspace{-0.1cm}

Measurements of the characteristics of \jpsi and \psip meson production in p-Pb collisions at 
$\rm \sqrt( s_{NN} ) $ have been performed. The differential cross sections for production of these particles in prompt and non-prompt modes have been determined and employed to characterize the dynamics of the interaction with the medium, including the forward-backward ratio and the non-prompt fraction~\cite{Arratia2015}. Further, differential cross sections for production of these mesons in p-p collisions at $\rm \sqrt(s)$ = 2.76 TeV have been measured. An interpolation of these cross sections, together with measurements from ATLAS for p-p collisions at $\rm \sqrt(s)$ = 7 and 8 TeV, has been performed in order to compare the p-Pb collision results for these mesons with p-p collisions at the same energy of $\rm \sqrt(s)$ = 5.02 TeV. The comparison of \jpsi and \psip meson production in p-Pb with p-p has been performed through calculation of $\rm R_{pPb}$ for prompt and non-prompt \jpsi and for prompt \psip as a function of $\rm p_T$ and of $\rm y^*$. These values are consistent with being greater than unity and approximately constant within the uncertainties. The dependence of the yields on event centrality has been investigated using several different centrality measures, including self-normalized ratios, normalization to Z bosons, and a double ratio described below. Both prompt and non-prompt \jpsi are consistent with being independent of event centrality, while a pattern of moderately decreasing yield of \psip with increasing centrality is apparent. The double ratio $\rm R_{pPb}( \Psi ( 2S ) ) /R_{pPb}( J/\Psi)$ has been measured, comparing \psip with \jpsi, and this ratio is equal to unity in the limit of no medium effects. This double ratio suggests enhancement for more-peripheral events, tending toward a possible suppression for more-central events, and the shape of the trend is very similar to that of the Z-normalized centrality dependence for \psip. These measurements represent a substantial step forward in understanding the dynamics of \jpsi and \psip production in systems involving the cold nuclear medium in QCD, as well as improving our understanding of centrality measures in p-Pb collisions. 







\end{document}